%% file: CameraReady.tex
\documentclass[10pt, conference]{IEEEtran}  % Comment this line out if you need a4paper
\usepackage{tikz}
\usepackage{acronym}
\usetikzlibrary{shapes,snakes}
\usepackage{empheq}
\usepackage{graphicx}
\usepackage{epstopdf}
\usepackage{multirow}
\usepackage{tabu}
\usepackage{array}
\usepackage{url}
\usepackage{amsmath}
\usepackage{cleveref}
\usepackage{amssymb}
\usepackage{amsthm}
\usepackage{epstopdf}
\usepackage{amsfonts}
\usepackage{tikz}
\usepackage{bm}
\usepackage{algorithmic}
\usepackage[ruled,vlined,linesnumbered]{algorithm2e}
\usepackage{graphicx}
\usetikzlibrary{arrows}
\usepackage{cite}
\usepackage{caption}
\usepackage{color}
\usepackage{subcaption}
\usepackage{amssymb}
\usepackage{tabulary}
\usepackage{booktabs}

\usepackage[justification=centering]{caption} % in order to centralize the caption in the figures
\DeclareCaptionLabelFormat{algnonumber}{algorithm}
\tikzstyle{int}=[draw, fill=white!20, minimum size=2em]
\tikzstyle{init} = [pin edge={to-,thin,black}]

\newcommand\blfootnote[1]{%
	\begingroup
	\renewcommand\thefootnote{}\footnote{#1}%
	\addtocounter{footnote}{-1}%
	\endgroup
}

 \DeclareRobustCommand*{\IEEEauthorrefmark}[1]{%
 	\raisebox{0pt}[0pt][0pt]{\textsuperscript{\footnotesize\ensuremath{#1}}}}

\input{acronyms}

\IEEEoverridecommandlockouts                          

\title{
	Optimal Sampling Cost in Wireless Networks with Age of Information Constraints
}

\author{
	\IEEEauthorblockN{Emmanouil Fountoulakis\IEEEauthorrefmark{\dagger},
		Nikolaos Pappas\IEEEauthorrefmark{\dagger},
		Marian Codreanu\IEEEauthorrefmark{\dagger},
		Anthony Ephremides\IEEEauthorrefmark{\dagger *}
		\IEEEauthorblockA{\IEEEauthorrefmark{\dagger} Department of Science and Technology, Link{\"o}ping University, Campus Norrk\"oping, Sweden}
		\IEEEauthorblockA{\IEEEauthorrefmark{*} Electrical and Computer Engineering Department, University of Maryland, College Park, USA}
		E-mails: \{emmanouil.fountoulakis, nikolaos.pappas, ionel.marian.codreanu\}@liu.se, etony@umd.edu}
}

\begin{document}
\immediate\write18{echo $PATH > tmp1}
\immediate\write18{/Library/TeX/texbin/epstopdf > tmp2} 

\maketitle
\thispagestyle{empty}
\pagestyle{empty}
	
\begin{abstract}
	We consider the problem of minimizing the time average cost of sampling and transmitting status updates by users over a wireless channel subject to average Age of Information constraints (AoI). 
	 Errors in the transmission may occur and the scheduling algorithm has to decide if the users sample a new packet or attempt for retransmission of the packet sampled previously. The cost consists of both sampling and transmission costs. The sampling of a new packet after a failure imposes an additional cost in the system. We formulate a stochastic optimization problem with time average cost in the objective under time average AoI constraints. To solve this problem, we apply tools from Lyapunov optimization theory and develop a dynamic algorithm that takes decisions in a slot-by-slot basis. The algorithm decides if a user: a) samples a new packet, b) transmits the old one, c) remains silent. We provide optimality guarantees of the algorithm and study its performance in terms of time average cost and AoI through simulation results.
\end{abstract}

%keywords
%\begin{IEEEkeywords}
%
%\end{IEEEkeywords}
	
\IEEEpeerreviewmaketitle 
	
\section{Introduction}
	\blfootnote{\begin{scriptsize}
		\noindent This work was supported in part by the Center for Industrial
		Information Technology (CENIIT), ELLIIT, and the Swedish Research
		Council (VR).
\end{scriptsize}}
The \ac{AoI} is a new metric that captures the timeliness or freshness of the data \cite{kostamonograph2017age, sunmodiano2019age}. It was first introduced in \cite{kaulyates2012real}, and it is defined as the time elapsed since the generation of the status update that was most recently received by a destination.  \ac{AoI} can play an important role in applications with freshness-sensitive data, e.g., environment monitoring, smart agriculture, sensor networks, etc. Consider a cyber-physical system, where a number of sensors sample and transmit freshness-sensitive data (e.g., temperature, humidity, solar radiation level) to a destination over a wireless channel. Under ideal conditions, the destination receives fresh data, continuously. However, due to the fluctuating nature of the channels and the limited resources, this is often impractical. In such cases, it is vital for the system to manage the resources efficiently to keep the data fresh. In this paper, we consider both the sampling and the transmission costs and propose a dynamic low-complexity algorithm that minimizes the time average cost of the system while keeping the time average \ac{AoI} below a threshold.

Recently, the performance analysis in terms of \ac{AoI} in queueing systems has attracted a lot of attention. In \cite{kamephr2015effect}, the authors consider a source that randomly generates packets  and transmits them to a remote monitor over a network with dynamic routes. The authors provide the approach for  computing the analytical status age  under different queueing models such as $M/M/1$, $M/M/2$, and $M/M/\infty$. In \cite{kostanonlinear2017age}, the authors introduce the notion of cost of update delay, and the value of information. By considering queue management in an $M/M/1$ system, the authors in \cite{kosta2019queuemanage} provide analytical results for the age of information and peak age of information of the system. Also, recent studies consider the application of \ac{AoI} in \ac{IoT} \cite{shreedhar2019age, corneo2019age, abd2019role, StamatakisIoT} systems or systems with energy harvesting capabilities \cite{chen2019optimal,bacinoglu2018achieving}.

The age minimization problem under different network scenarios is considered in \cite{kadota2019scheduling,kadota2019minimizing,hsu2017age}. In \cite{kadota2019scheduling}, the authors consider the \ac{AoI} minimization problem in a system with users that sample fresh information and transmit it to a destination over a wireless channel. In \cite{kadota2019minimizing}, the authors consider the minimization of \ac{AoI} problem in a system which packets  randomly arrive in a base station. The packets  are enqueued in separate queues and they are transmitted to the corresponding destinations for keeping the information fresh. The authors in \cite{hsu2017age} formulate the age minimization problem as a Markov Decision Process and provide the optimal on-line and off-line scheduling algorithms. In \cite{moltafet2019power}, the authors consider the power minimization problem in a wireless network with average \ac{AoI} constraints. The authors apply tools from Lyapunov optimization theory and provide an algorithm that minimizes the power consumption while keeping the data fresh in the destination. In an \ac{AoI}-constrained wireless network, authors in \cite{he2019emptying} consider the link activation problem for energy minimization. \textit{To the best of our knowledge there is no work that considers both transmission and sampling costs in a wireless network with time average \ac{AoI} constraints.}

In this paper, we formulate the time average cost minimization problem of an \ac{AoI}-constrained system. The cost consists of both transmission and sampling costs. Due to the probable failures of transmissions, the following question arises: should the users sample or attempt for re-transmission of the old packet? To address this question, we formulate a stochastic optimization problem for minimizing the time average cost under time average \ac{AoI} constraints. We apply tools from Lyapunov optimization theory in order to solve this problem and provide a dynamic low-complexity algorithm. We prove that the algorithm provides arbitrarily close to the optimal solution. In addition, we analyze the performance of the algorithm in terms of time average cost and \ac{AoI} through simulation results.

\section{System Model}\label{Sec:SytemModel}
%\begin{figure}
%	\centering
%	\includegraphics[scale=0.2]{Image/Fig1}
%	\caption{System model.}
%	\label{fig:systemmodel}
%\end{figure}

We consider a set of users $\mathcal{N}=\{1,\ldots,N\}$ who sample fresh information and send this information, in form of packets, to a receiver over a wireless fading channel. Time is assumed to be slotted, let $t$ $\in$  $\mathbb{Z}_{+}$ be the $t^{\text{th}}$ slot. We consider a scheduler that decides at every time slot the sampling and transmission scheduling of the users. At every time slot, each user is either decided to sample fresh information and send it to the receiver, or to transmit an old packet that has been sampled previously, or to remain silent. We denote by $s_{i}(t)$ the decision of user $i$ to sample  in time slot $t$, where   
\begin{align}\label{eq: samplingdecision}
s_{i}(t)=
\begin{cases}
1, \text{if user $i$ is decided to sample,}\\
0,\text{otherwise.}
\end{cases}
\end{align}
Note that each user samples the information at the beginning of the slot, if it is decided, and transmits the packet to the receiver by the end of the same slot. We denote by $\mu_{i}(t)$ the decision of user $i$ to transmit in the $t^{\text{th}}$ slot where 
\begin{align}\label{eq:transmdecision}
\mu_{i}(t)=
\begin{cases}
 1, \text{ if user } i \text{ is decided to transmit,} \\
 0\text{, otherwise}.			
\end{cases}
\end{align}
Note that if user $i$ samples new information in the $t^{\text{th}}$ slot, it attempts for transmission in the same slot. Therefore, if $s_{i}(t)=1$ then, $\mu_{i}(t)=1\text{.}$ However, in the case which  user $i$ does not sample new information but it has an old packet, variable $\mu_{i}(t)$ can take either the value of $0$ or $1$.
Furthermore, we assume that the users always transmit with fixed power transmission. Since we consider fading channels, errors in the transmission may occur. We assume that a packet is successfully transmitted from user $i$ to the receiver with a probability $p_{i}$\footnote{$p_{i}$ captures fading and noise in a wireless channel.}. The success  probability remains the same over the time, but it can be different from one user to another. We assume that the receiver sends an instantaneous ACK for a successful packet reception. We impose that we can have up to one user transmitting in one slot which is described by the following constraint
\begin{align}\label{constr: interference}
	\sum\limits_{i=1}^{N} \mu_{i}(t)\leq 1\text{, } \forall t\text{.}
\end{align}
At each time slot we can have up to one packet reception in the receiver.
We denote by $d_{i}(t)$ the successful packet reception from user $i$ to the receiver, where
\begin{align}
	d_{i}(t)=
	\begin{cases}
		1, \text{if user i transmits successfully a packet}\text{,}\\
		0, \text{otherwise.}
	\end{cases}
\end{align}
If  $\mu_{i}(t)=1$, then $d_{i}(t)$ takes the value of one with probability $p_{i}$ and zero with probability $1-p_{i}$. On the other hand, if $\mu_{i}=0$, then $d_{i}(t)$ takes the value of zero with probability one. Note that there are two cases which $\mu_{i}(t)$ can take the value of one; a) when user $i$ is decided to sample and transmit, b) when user $i$ is decided to transmit an old packet. It follows that $\mathbb{E}\{d_{i}(t) | \mu_{i}(t), s_{i}(t)\} = p_{i}\mu_{i}(t) + p_{i}s_{i}(t) -p_{i}s_{i}(t)\mu_{i}(t)\text{.}$ By applying the law of iterated expectations, we obtain
\begin{align}\label{expectedd_i}
	\mathbb{E}\{d_{i}(t) \} &= p_{i}\mathbb{E}\{\mu_{i}(t)\} + p_{i}\mathbb{E}\{s_{i}(t)\}
	 -p_{i}\mathbb\mathbb{E}\{s_{i}(t)\mu_{i}(t)\}\text{.}
\end{align}

\subsection{Age of Information}
The Age of Information (AoI) represents how ``fresh" is the information from the perspective of the receiver. Let $A_{i}(t)$ be a strictly positive integer that depicts the AoI associated with user $i$ at the receiver. If the received packet has been sampled at the beginning of the current slot, then $A_{i}(t+1)=1$. On the other hand, if the received packet has been sampled in a previous slot, then the age of information depends also in the time of the packet waiting for successful transmission. In order to characterize the waiting time of a packet in user $i$, we first define the sampling time, i.e., the time slot that the packet has been sampled.
We denote the last sampling time by $t^s$. Thus, the age of the packet that is in user $i$ is 
\begin{align}
	A^{p}_i(t) = t-t^{s}\text{, }\forall i \in \mathcal{N}\text{,}
\end{align}
where $t$ is the current slot. Therefore, the evolution of the total age of information at the receiver with respect to the user $i$ is written as
\begin{align}
 A_{i}(t+1) = 
 \begin{cases}
  A_i^p(t)+1\text{, if } d_{i}(t)=1\text{,}\\
  A_{i}(t) +1\text{, otherwise.}
 \end{cases}
\end{align}
Note that if a packet is successfully transmitted in the same slot which has been sampled, then $A^{p}_{i}(t)=0$ and therefore, $A_{i}(t+1)=1$. The evolution of \ac{AoI} can be written  as
\begin{align}\label{eq: ageevolutioncompact}
	A_{i}(t+1) = (A_{i}^p+1)d_{i}(t) + (A_{i}(t)+1)(1-d_{i}(t))\text{.}
\end{align}

Furthermore, for each sampling or transmission, we consider an associated cost for each user. We denote by $c_s$ and $c_\text{tr}$ the cost for sampling and transmission for each user, respectively. We consider that the costs remain constants over time.  The cost function of user $i$ in a time slot is described below
\begin{align}
 c_{i}(t) = \mu_{i}(t)c_{\text{tr}} + s_{i}(t)c_s\text{, } \forall i \in \mathcal{N}\text{.}
\end{align}
The total cost of the system is described as
\begin{align}
	c(t) = \sum\limits_{i=1}^N c_{i}(t)\text{,}
\end{align}
and the expected time average cost as 
\begin{align}
	\bar{c} = \lim_{t\rightarrow \infty}\frac{1}{t}\sum\limits_{\tau=0}^{t} \mathbb{E}\{c(\tau)\}\text{.}
\end{align}
The  expected time average age for user $i$ is described as
\begin{align}
	\bar{A_i} = \lim_{t\rightarrow \infty} \frac{1}{t}\sum\limits_{\tau=0}^t\mathbb{E}\{A_i(\tau)\}\text{, } \forall i \in \mathcal{N}\text{,}
\end{align}
where the expectations are with respect to the channel randomness and the scheduling policy. 
Let $A_{i}^{\text{max}}$ be a strictly positive real value that represents the maximum expected time average age requirement of user $i$, and it is described by the constraint:
\begin{align}\label{const: age}
 \bar{A_i}\leq A_i^{\text{max}}\text{, } \forall i \in \mathcal{N}\text{.}
\end{align}
%In this paper, we consider that there exists policy $\pi$ $\in$ $\Pi$ that satisfies all interference constraints in $\eqref{constr: interference}$ and maximum time expected average age constraints in $\eqref{const: age}$, simultaneously.

\subsection{Optimization Problem}
With the definitions of AoI and expected time average costs, we define the stochastic optimization problem as following. 
\begin{subequations}
	\begin{align}\label{Opt.Probl}
	\min\limits_{\bm{s}(t)\text{, } \bm{\mu}(t) } \quad & \bar{c}\\\label{constropt: inteference}
	\text{s.~t.} \quad & \sum\limits_{i=1}^{K} \mu_{i}(t) \leq 1\text{, } \forall t\text{,}\\\label{constropt: age}
	\quad & \bar{A}_{i}\leq A_{i}^{\text{max}}\text{, } \forall i \in \mathcal{N}\text{,}\\
	\quad & \bm{s}(t)\text{, } \bm{\mu}(t) \in \{0,1\}^{N}\text{,} 
	\end{align}
\end{subequations}
where $\bm{\mu}(t)= [\mu_{1}(t),\ldots,\mu_{N}(t)]$ and $\bm{s}(t) = [s_{1}(t),\ldots,s_{N}(t)]$.
The interference and maximum expected time average age constraints are depicted in \eqref{constropt: age} and \eqref{constropt: inteference}, respectively. Our target is to find a policy, with optimality guarantees, that minimizes the system cost while providing time average age below a threshold for each user $i$.
\subsection{Proposed  Solution}
In this section, we provide a low-complexity scheduling algorithm that satisfies the time average age constraints and provides solution arbitrarily close to the optimal one.

We apply the technique, first developed in  \cite{TassiulasNeelyNow} and further discussed in \cite{neely2010stochastic} and \cite{NeelyEnergyOptimalControl}, in order to satisfy the time average age constrains in \eqref{constropt: age}. Each inequality constraint is mapped into a virtual queue. We show below that the time average age problem is transformed into a queue stability problem. 

Let $\{X_{i}(t)\}_{i \in \mathcal{N}}$ be the virtual queues associated with constraints in $\eqref{constropt: age}$. We update each virtual queue $i$ at each time slot $t$ as
\begin{align}\label{vqevolution}
 X_{i}(t+1) = \max [X_{i}(t)-A_{i}^{\text{max}}, 0] + A_{i}(t+1)\text{.}
\end{align}
Process $X_{i}(t)$ can be viewed as a queue with ``arrivals"  $A_{i}(t)$ and service rate $A_{i}^{\text{max}}$.
Before describing the details of the analysis, let us recall a basic theorem that is based on the general theory of the stochastic processes \cite{MeynMarkovChains}. Consider a system with $K$ queues. The number of unfinished jobs of queue $k$ is denoted by $q_{i}(t)$ and $\mathbf{q}(t) = \{q_{k}(t)\}_{k \in \mathcal{K}}$. The Lyapunov function and the the Lyapunov drift are denoted by $L(\mathbf{q}(t))$ and $\Delta(L(\mathbf{q}(t))) \triangleq \mathbb{E}\{L(\mathbf{q}(t+1)-L(\mathbf{q}(t)))|\mathbf{q}(t)\}$, respectively.\\
\noindent \textit{Definition 1 (Lyapunov Function)}: A function $L: \mathbb{R}^{K} \rightarrow \mathbb{R}$ is said to be a Lyapunov function if it has the following properties
\begin{enumerate}
	\item It is non-decreasing in any of its arguments.
	\item $L(\mathbf{x}) \geq 0\text{, } \forall \mathbf{x} \in \mathbb{R}^{K}$\text{.}
	\item $L(\mathbf{x}) \rightarrow + \infty$, as $\|\mathbf{x}\| \rightarrow + \infty$.
\end{enumerate}
\textit{Theorem 1. (Lyapunov Drift): If there are positive values B, $\epsilon$ such that for all time slots $t$ we have $\Delta(L(\mathbf{q}(t))) \leq B - \epsilon \sum\limits_{k=1}^K q_n (t)$, then the system $\mathbf{q}(t)$ is strongly stable.}
\subsection{Drift-Plus-Penalty Policy}
The \ac{DPP} algorithm is designed to minimize the sum of the Lyapunov drift and a penalty function \cite[Chapter 3]{neely2010stochastic}. First, we define the Lyapunov drift as
\begin{align}\label{eq: lyapunovdrift}
	\Delta (\mathbf{X}(t)) = \mathbb{E} \{L(\mathbf{X}(t+1)) - L(\mathbf{X}(t)) | S_t \} \text{,}
\end{align}
where $S_t= \{A_i(t), X_{i}(t)\}_{i \in \mathcal{N}}$ is the network state at the beginning of slot $t$ and $\mathbf{X}(t) = \{X_i(t)\}_{i \in \mathcal{N}}$. The associated Lyapunov function is defined as 
\begin{align}  
	L = \frac{1}{2} \sum\limits_{i=1}^{N} X^2_{i}(t)\text{.}
\end{align}
The above expectations are with respect to the channel randomness and the scheduling policy. We apply the \ac{DPP} algorithm to minimize the time average cost while stabilizing the virtual queues $\{X_{i}(t)\}_{i \in \mathcal{N}}$. In particular, this approach seeks to minimize an upper bound of the following expression
\begin{align}\label{expr: driftpluspenalty}
			\Delta (\mathbf{X}(t))  + V \mathbb{E}\{c(t) |S_{t}\}\text{,}
\end{align}
where $V$ is an importance weight to scale the penalty. An upper bound for the expression in \eqref{expr: driftpluspenalty} is shown below 
\begin{align}\nonumber
  &	\Delta (\mathbf{X}(t))  +  V \mathbb{E}\{c(t) | S_{t}\} \leq B + \sum\limits_{i=1}^N \mathbb{E} \{
  	X_{i}(t) [(A_i^p(t)+1 )W_{i}(t) \\ \label{ineq: dppbound}
  	&+  (A_{i}(t)+1) (1-W_{i}(t)) - A^\text{max}_i ] | S_t \} + V\mathbb{E}\{c(t)|S_{t}\}\text{,}
\end{align}
where $W_{i}(t) = p_i s_i(t) + p_i\mu_{i}(t) - p_i s_i(t)\mu_{i}(t)$, and $B\geq \sum\limits_{i=1}^N \frac{\mathbb{E}\{(A_{i}(t+1)^2|S_t\} + (A_{i}^{\text{max}})^2}{2}  $. The complete derivation of the above bound can be found in Appendix A.  For $B=\sum\limits_{i=1}^{N} \frac{(A_{i}(t)+1)^2 + (A_{i}^{\text{max}})^2}{2}$, we see that $B$ is not affected by the decisions $s(t), \mu(t)$ at each slot $t$. Therefore, we can exclude it from the optimization problem and minimize the second term of \eqref{ineq: dppbound}. The DPP algorithm takes sampling and transmission decisions at each time slot by solving the following optimization problem.
\begin{subequations}
	\begin{align}\nonumber
	\min\limits_{\bm{\mu}(t),\bm{s}(t)} \quad &   \sum\limits_{i=1}^N  \{
	X_{i}(t) [(A_i^p(t)+1 )W_{i}(t) \\ 
	&+  (A_{i}(t)+1) (1-W_{i}(t)) - A^\text{max}_i ] \} + Vc(t)\\
	\text{s.~t.} \quad & \sum\limits_{i=1}^{K} \mu_{i}(t)\leq 1\text{, } \forall t\text{,}\\
	\quad & \bm{s}(t)\text{, } \bm{\mu}(t) \in \{0,1\}^{N} \text{.} 
	\end{align}
\end{subequations}
\textit{Theorem 2. (Optimality of the DPP algorithm and virtual queue stability): The DPP algorithm guarantees that the virtual queues are strongly stable and therefore, the time average age constraints in \eqref{constropt: age} are satisfied. In particular, the time average expected value of $X_{i}(t)$ is bounded as
\begin{align}\label{bound: virtqueue}
\lim_{t\rightarrow \infty} \sup \frac{1}{t} \sum\limits_{\tau=0}^{t}\sum\limits_{i=1}^{N} \mathbb{E} \{X_{i}(\tau)\} \leq \frac{B+V(c^{*}(\epsilon) - c^{\text{opt}})}{\epsilon}\text{.}
\end{align}	
 In addition, the expected time average cost is bounded as
	\begin{align}\label{bound: cost}
	  \lim_{t\rightarrow \infty} \sup \frac{1}{t} \sum\limits_{\tau=0}^{t-1} \mathbb{E} \{c(\tau)\} \leq c^{\text{opt}} +\frac{B}{V}\text{.}
	\end{align}
}
\begin{proof}
	See Appendix B.
\end{proof}
\textbf{Remark 1. }Theorem 1  indicates that the DPP algorithm provides a solution arbitrarily close to the optimal one. We can get better performance in terms of time average cost by increasing the value of $V$. However, we observe from \eqref{bound: virtqueue} that the time average age increases as $V$ increases. Therefore, there is a trade-off between the  time average cost and time average age.

\section{Simulation Results}
In this section, we provide results in order to evaluate the performance of our algorithm in terms of average $\ac{AoI}$ and cost of system. We consider a system with two users that transmit fresh information to the receiver over an unreliable wireless channel. First, we provide results to observe the effect of the importance factor $V$ on the average \ac{AoI} and cost. In addition, we observe the behavior of the scheduler for the case where one user has lower success probability than the other. Second, we provide results for different values of the success probabilities and the effect on the average AoI and cost of the system.

\begin{figure}[t!]
	\begin{subfigure}{.5\textwidth}
		\centering
	\includegraphics[scale=0.53]{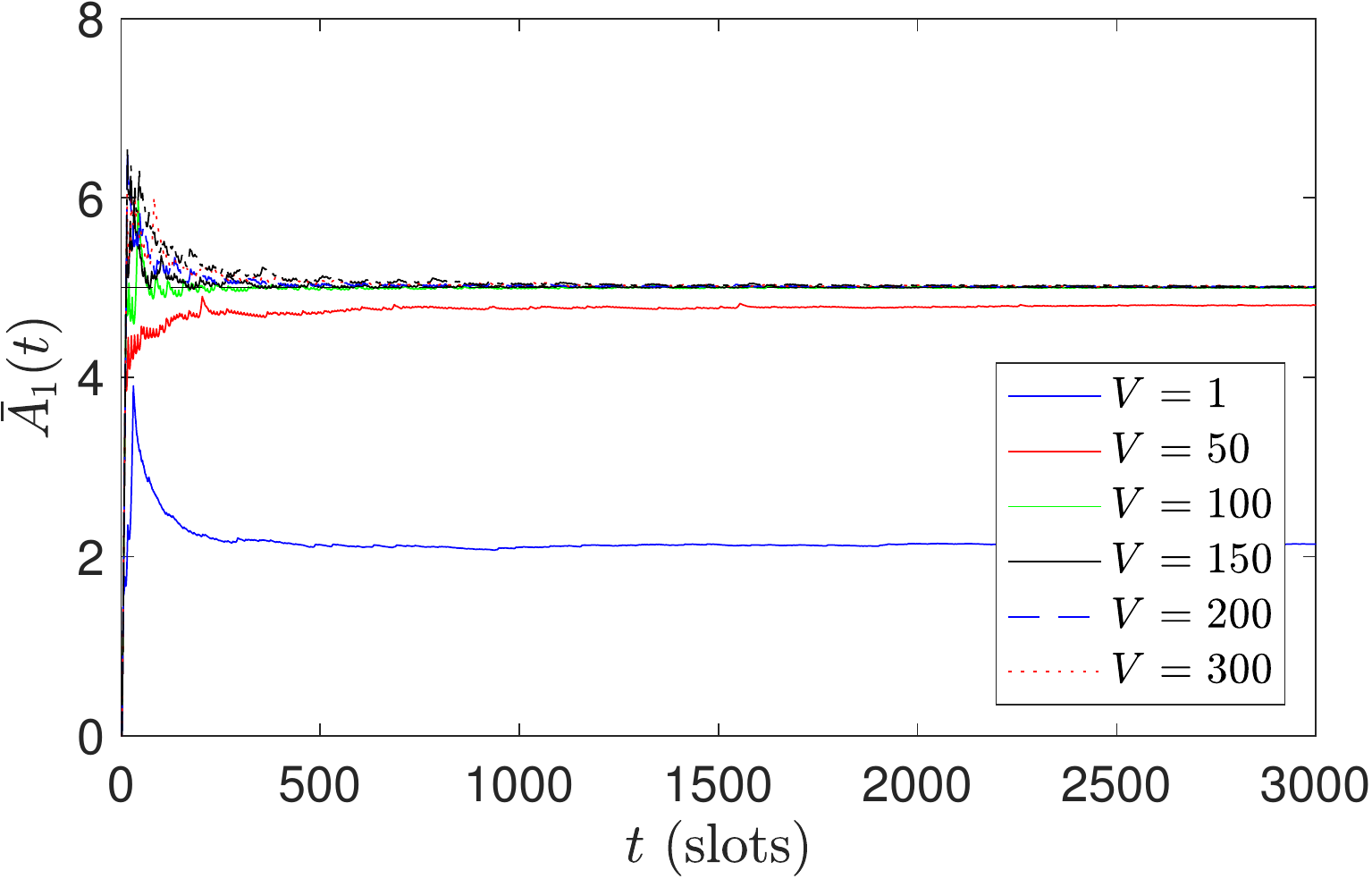}
	\caption{Average AoI of user $1$ for different values of $V$. }
	\label{fig:averageageslotsuser1diffprob}
	\end{subfigure}%
	
	\centering
	\begin{subfigure}{.5\textwidth}
		\centering
	\includegraphics[scale=0.53]{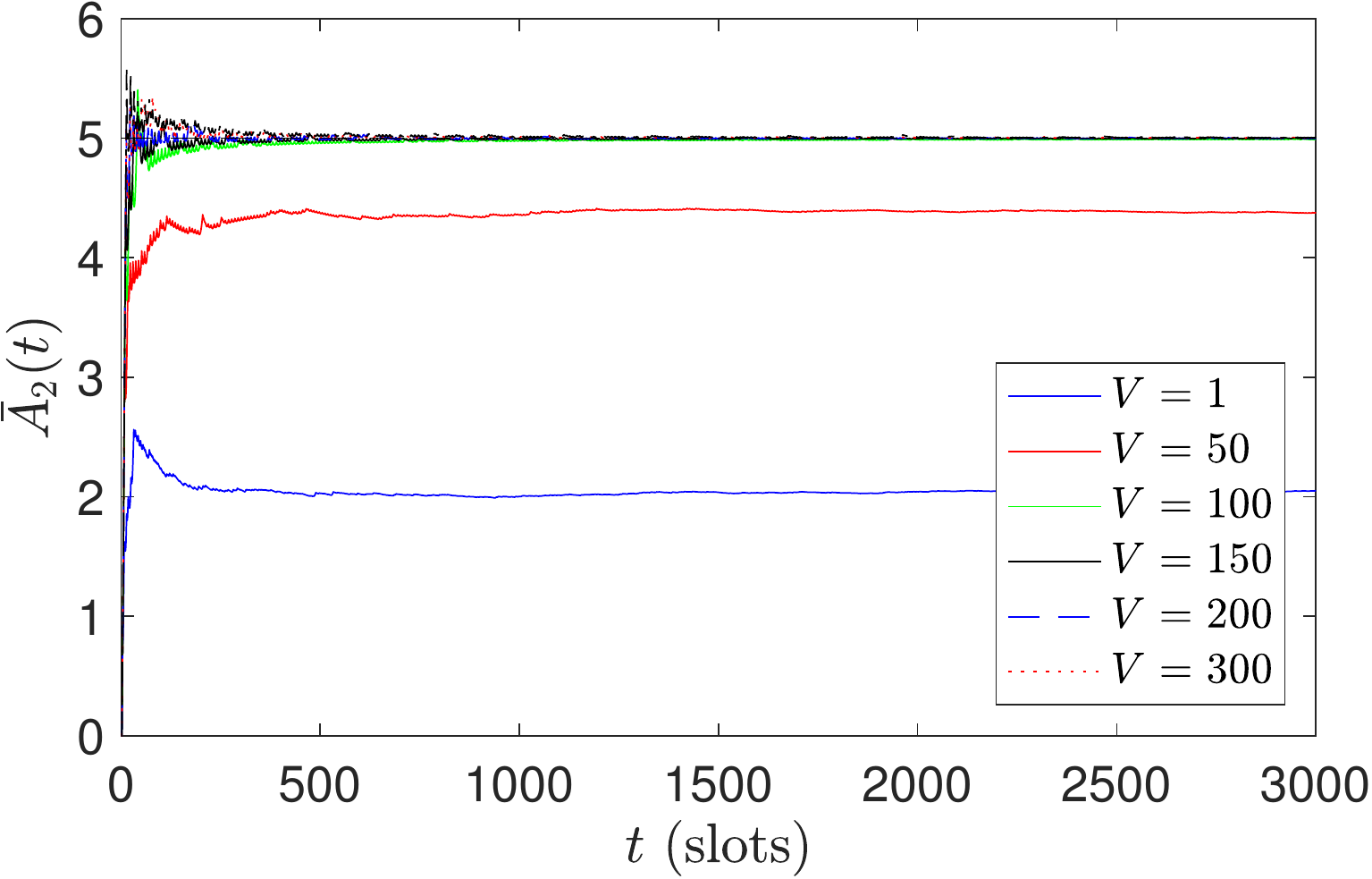}
	\caption{Average AoI of user $2$ for different values of $V$.  }
	\label{fig:averageageslotsuser2diffprob}
	\end{subfigure}
	
	\centering
	\begin{subfigure}{.44\textwidth}
		\centering
	\includegraphics[scale=0.53]{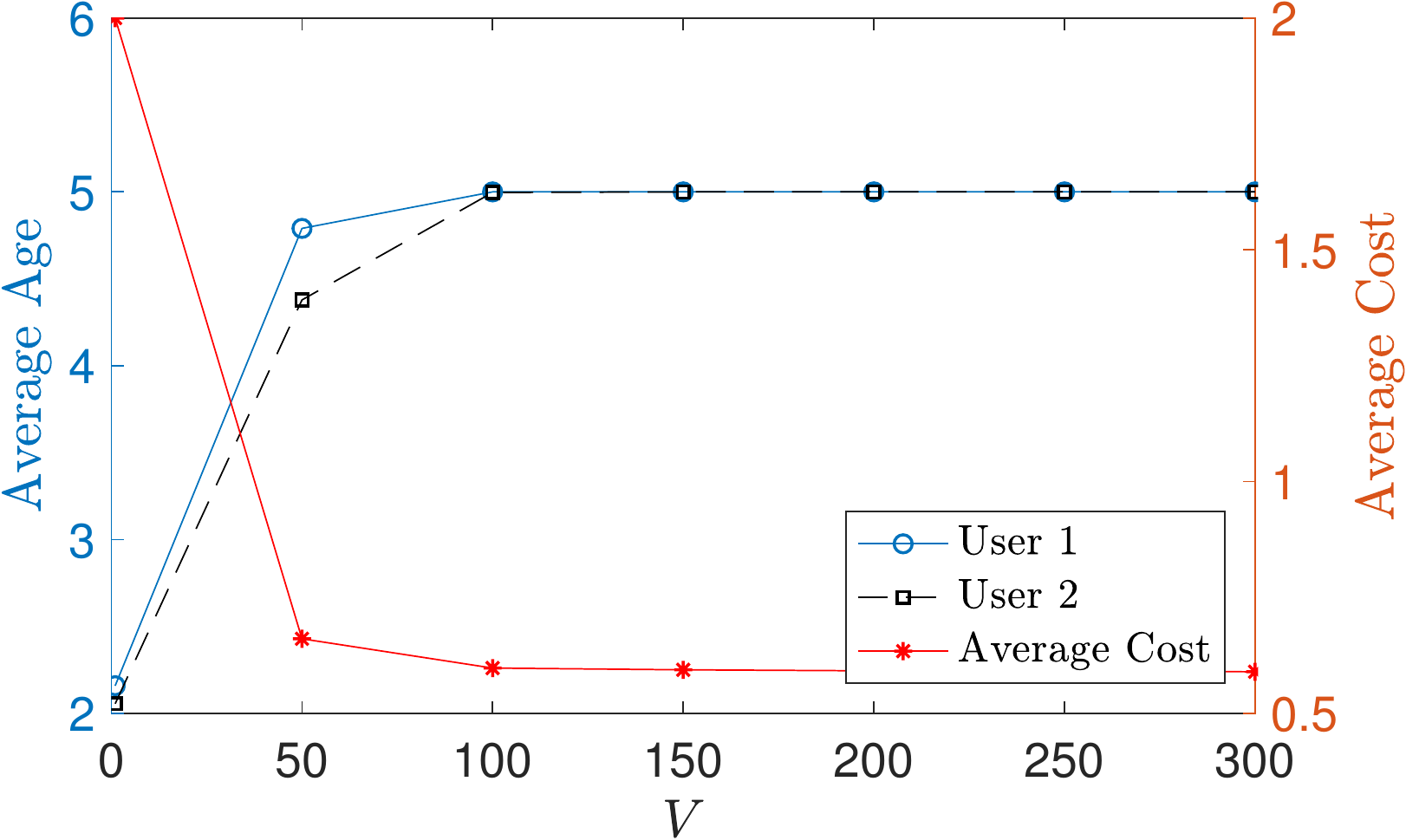}
	\caption{Average cost vs average AoI. }
	\label{fig:averageagevscostdiffprob}
	\end{subfigure}
	\caption{System with $2$ users. $p_{1}=0.6$, $p_{2}=0.9$, $A_{1}^{\text{max}}=A_{2}^{\text{max}}=5$.}
	\label{Fig:TradeAgeEvolutionCostAverageAge}
\end{figure}

\begin{figure}[t!]
	\centering
	\includegraphics[scale=0.43]{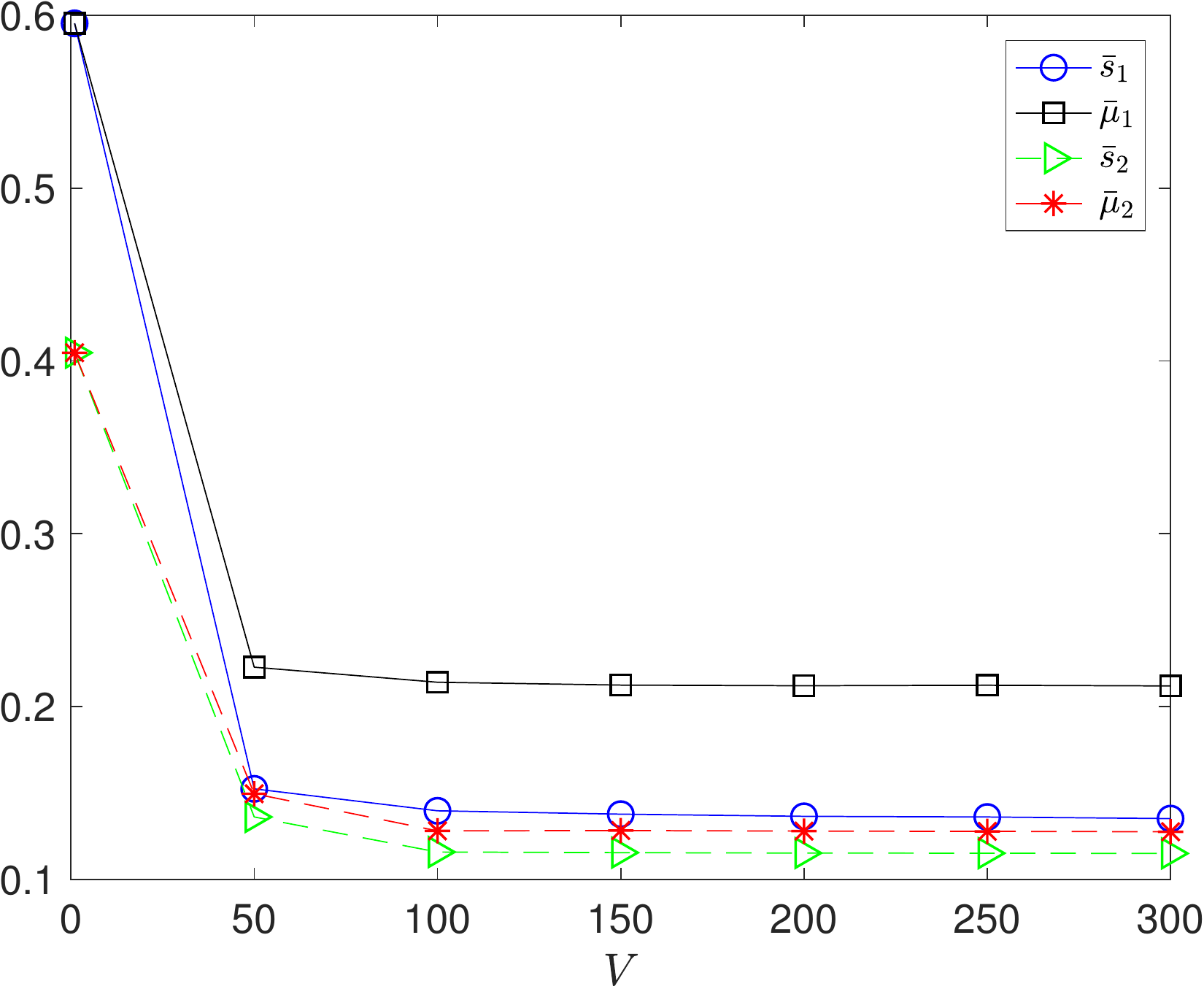}
	\caption{Average number of transmissions and average number of samplings. $p_{1}=0.6$, $p_{2}=0.9$, $A_{1}^{\text{max}}=A_{2}^{\text{max}}=5$.}
	\label{fig:vvsmusdiffprob}
\end{figure}

In Figs. \ref{Fig:TradeAgeEvolutionCostAverageAge} and \ref{fig:vvsmusdiffprob}, we provide results for a system that consists of  user $1$ and user $2$ with $p_{1}=0.6$ and $p_{2}=0.9$, respectively. Fig. \ref{Fig:TradeAgeEvolutionCostAverageAge} depicts the average AoI of user $1$ over time  for different values of the importance factor $V$. Recall that $V$ is a factor that is multiplied by the cost. Therefore, if our goal is to decrease the cost of the system we increase $V$. However, higher values of $V$ affect the average AoI of the system as shown in Fig. \ref{fig:averageageslotsuser1diffprob}. We observe that as $V$ increases, the convergence time of the algorithm increases as well. For example, for $V=50$, we observe that $\bar{A}_{1}(t)$ takes values less than $A_{1}^{\text{max}}$ during the first slots. On the other hand, for $V=300$, we observe that $\bar{A}_{1}(t)$ takes values less than $A_{1}^{\text{max}}$ after $1500$ slots. In Fig. \ref{fig:averageagevscostdiffprob}, we observe that average AoI for different values of $V$. We observe that small values of $V$, i.e. values between $1$ and $100$, user $1$ has a higher average AoI because of the smaller success probability. Furthermore, we verify our theoretical results regarding the time aveage AoI constraints. 

In addition, we observe from Fig. \ref{fig:averageageslotsuser1diffprob} and \ref{fig:averageageslotsuser2diffprob} the effect of value of the success transmission probability on the average AoI and its evolution over time. As the success probability decreases the average AoI increases because we need more time to accomplish a successful transmission. Therefore, the packet arrives after some attempts to the receiver from user $1$, thus, AoI increases.


 \begin{figure}[h!]
	\centering
	\includegraphics[scale=0.43]{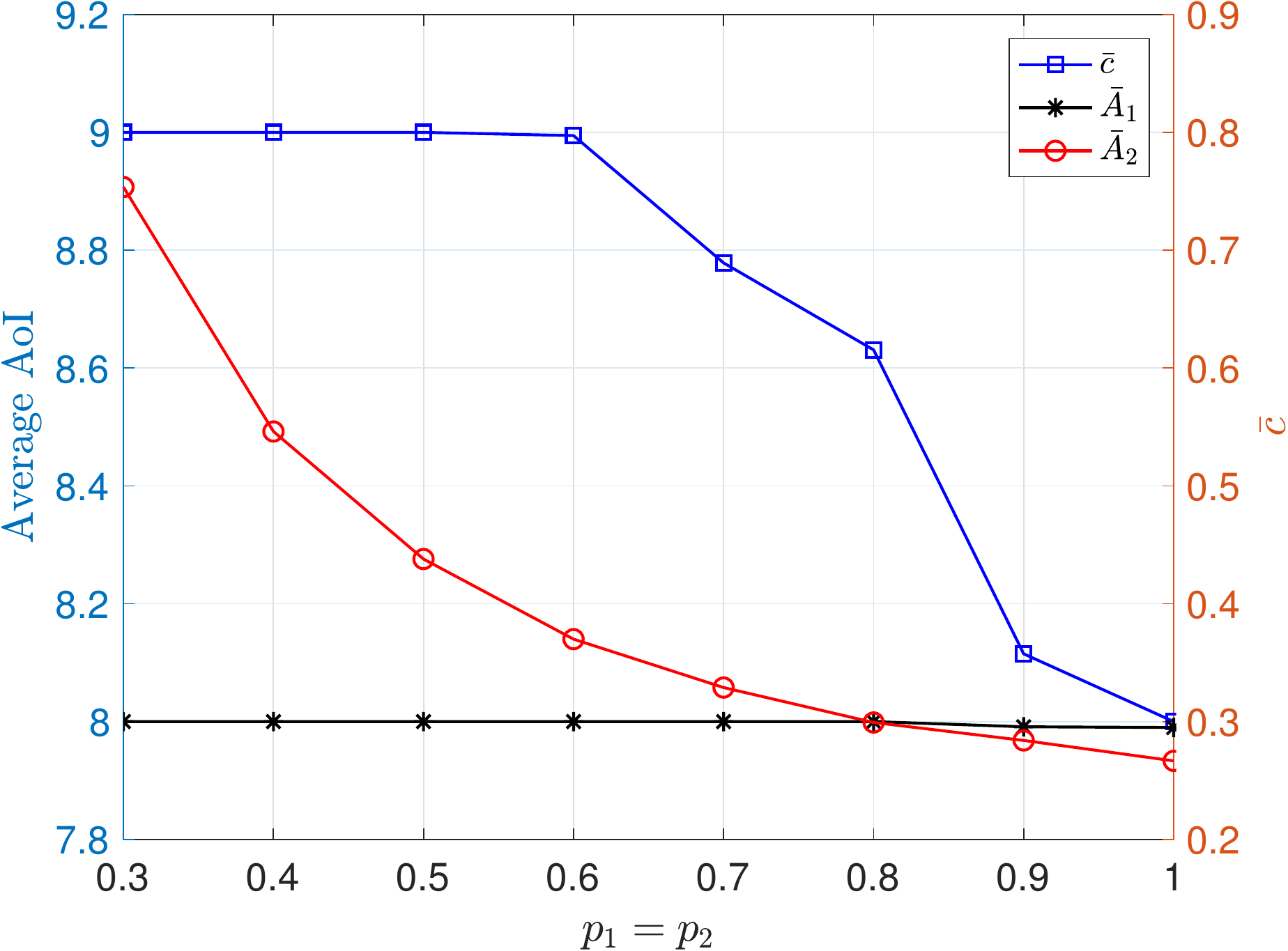}
	\caption{Average \ac{AoI} for user $1$ and user $2$  and the total cost for different values of $p_1$ and $p_2$. $A_{1}^{\text{max}}=9$ and $A_{2}^{\text{max}}=8$. $V=200$.}
	\label{fig:CostAgeVarPRob}
\end{figure}
In Fig. \ref{fig:CostAgeVarPRob}, we observe the average \ac{AoI} for user $1$ and user $2$ and the cost for different values of the success probabilities, $p_1$, $p_2$. In this setup, we have $A_{1}^{\text{max}}=8$ and $A_{2}^{\text{max}}=9$. We observe that as the success probability increases, it is less costly for the system to satisfy the AoI constraints for both users. Therefore, the algorithm utilizes efficiently the resources of the system. Furthermore, we observe that for large values of success transmissions, the average AoI of user $1$ decreases. Therefore, the algorithm utilizes the resources such that not even to satisfy the constraints but also to provide a better performance in terms of average AoI. The constraints for user $2$ are also satisfied. Note that the average AoI constraints of user $1$ are harder than 
user $2$. This does not allow the algorithm to give a better performance for user $1$  without increasing the cost and therefore, the scheduler selects to satisfy the constraints with the lowest possible cost.
%\subsection{Two user with identical success probabilities}
%\begin{figure}[h!]
%	\centering
%	\includegraphics[scale=0.4]{Figures/AverageAgeSlotsUser1EqProb}
%	\caption{Average AoI of user $1$ for different values of $V$. $p_{1}=p_2=0.9$. $A_{1}^{\text{max}}=A_{2}^{\text{max}}=5$.}
%	\label{fig:averageageslotsuser1eqprob}
%\end{figure}
%\begin{figure}[h!]
%	\centering
%	\includegraphics[scale=0.4]{Figures/AverageAgeSlotsUser2EqProb}
%	\caption{Average AoI of user $2$ for different values of $V$. $p_{1}=p_2=0.9$. $A_{1}^{\text{max}}=A_{2}^{\text{max}}=5$.}
%	\label{fig:averageageslotsuser2eqprob}
%\end{figure}
%\begin{figure}[h!]
%	\centering
%	\includegraphics[scale=0.4]{Figures/AverageAgeVsCostEqProb}
%	\caption{Average cost vs average AoI. $p_{1}=p_{2}=0.9$, $A_{1}^{\text{max}}=A_{2}^{\text{max}}=5$.}
%	\label{fig:averageagevscosteqprob}
%\end{figure}
%\begin{figure}[h!]
%	\centering
%	\includegraphics[scale=0.4]{Figures/VvsMuSEqProb}
%	\caption{Average number of transmissions and average number of samplings. $p_{1}=p_{2}=0.9$, $A_{1}^{\text{max}}=A_{2}^{\text{max}}=5$.}
%	\label{fig:vvsmuseqprob}
%\end{figure}
%
%\begin{figure}
%	\centering
%	\includegraphics[scale=0.4]{Figures/CostHarndnessofAge}
%	\caption{Time average cost vs hardness of AoI constraints. $p_1=p_2=0.9$.}
%	\label{fig:costharndnessofage}
%\end{figure}

\section{Conclusions}
In this paper, we propose an algorithm that decides the sampling and transmission scheduling at each time slot by minimizing the time average cost of an AoI-constrained wireless system. The proposed algorithm is based on Lyapunov optimization theory. We prove that the algorithm can provide a solution arbitrarily close to the optimal. We consider cost for both the transmission and sampling. Due to the fading  channel, errors may occur and therefore, transmissions need to be optimized. We provide simulation results to evaluate the performance of our algorithm in terms of average AoI and cost. The results show the trade-off between the cost and average AoI and how different values of success probabilities affect the performance of the system.

\begin{appendices}
\section{Upper Bound on the Lyapunov Drift of DPP}
Using the fact that $\left( \max[Q-b,0]+A \right)^2 \leq Q^2 +A^2 +b^2+2Q(A-b)$, we rewrite  \eqref{vqevolution} as 
\begin{align}\nonumber
  X^2_{i} & \leq X_{i}^2(t) + A_{i}^2(t+1) +(A_{i}^{\text{max}})^2 \\\label{1st}
  & + 2X_{i}(t) (A_{i}(t+1)-(A_{i}^{\text{max}})^2)\text{.}
\end{align}
Rearranging the terms in \eqref{1st} and dividing by $2$, we take
\begin{align}\nonumber
		&\sum\limits_{i=1}^{N} X_{i}^2(t+1) - X^2_{i}(t)\\\label{2st}
		 &\leq \sum\limits_{i=1}^N \frac{A_{i}^{2}(t+1) + (A_{i}^{\text{max}})^2 + 2X_{i}(t)(A_{i}(t+1)-A_{i}^{\text{max}})}{2}\text{,}
\end{align}
 taking expectations in \eqref{2st}, we obtain
 \begin{align}\nonumber
 	\Delta(\mathbf{X}(t)) & \leq \sum\limits_{i=1}^{N} \frac{\mathbb{E}\{A_{i}^2(t+1)| S_t\} + (A_{i}^{\text{max}})^2}{2}\\\label{3nd}
 	& + \sum\limits_{i=1}^NX_{i}(t) [\mathbb{E}\{A_{i}(t+1) | S_t \}- A_{i}^{\text{max}}] \text{.}
 \end{align}
To obtain the expression associated with AoI, we calculate $\mathbb{E} \{A_{i}(t+1) | S_{i}(t))\}$ and  $\mathbb{E} \{A_{i}^2(t+1) | S_t\}$ using the evolution in \eqref{eq: ageevolutioncompact}. It follows that
\begin{align}\nonumber
	&\mathbb{E}\{A_{i}(t+1) | S_t\} =\\ \nonumber
	&	= \mathbb{E}\{d_{i}(t)(A_{i}^p(t)+1)|S_t\} 
	+ \mathbb{E}\{(1-d_{i}(t))(A_i(t)+1)|S_t\}\\\nonumber
	&	= \mathbb{E}\{(A_{i}^p(t)+1) (p_is_{i}(t) + p_{i}\mu_{i}(t)
	- p_{i}s_{i}(t)\mu_{i}(t))|S_t\}\\\label{ageboundsquare}
	& + \mathbb{E}\{(A_i(t)+1)(1- p_is_{i}(t) - p_{i}\mu_{i}(t)
	+ p_{i}s_{i}(t)\mu_{i})|S_t\}\text{,}
\end{align}
and 
\begin{align}\nonumber
			\mathbb{E}\{A_i^2(t+1)|S_t\} &= \mathbb{E}\{(A_{i}^p(t)+1)^2d_{i}(t) |S_t \} \\\label{agebound}&+\mathbb{E}\{(A_i(t)+1)^2(1-d_{i}(t))|S_t\}\text{.}
\end{align}

\noindent By applying \eqref{agebound} and \eqref{ageboundsquare} in \eqref{3nd}, we obtain
\begin{small}
\begin{align}\nonumber
	&\Delta(\mathbf{X}(t)) \leq \sum\limits_{i=1}^N\frac{\mathbb{E}\{(A^2_{i}(t+1))|S_t\}+(A_{i}^{\text{max}})^2}{2}\\\nonumber
	&+\sum\limits_{i=1}^N\mathbb{E}\{X_{i}(t)[(A_{i}^p(t)+1)\times(p_is_{i}(t)+ p_i\mu_{i}(t) - p_is_{i}\mu_{i})\\\nonumber
	&+(A_{i}(t)+1)(1-p_{i}s_{i} (t) 
	- p_i\mu_{i}(t)+p_{i}\mu_{i}(t)s_{i}(t))] \\\label{eq28}
	&- X_{i}(t)A_{i}^{\text{max}}|S_t\}\text{.}
\end{align}
\end{small}

\noindent We consider that there is a $B$ such that 
\begin{align}
		B\geq \sum\limits_{i=1}^{N} \frac{\mathbb{E}\{A^2_{i}(t+1)|S_t\}+(A_{i}^\text{max})^2}{2}\text{.}
\end{align}
By setting $W_{i}(t)=p_is_i(t) + p_{i}\mu_{i}(t) - p_i\mu_{i}(t)s_{i}(t)$ in \eqref{eq28}, we obtain the result in \eqref{ineq: dppbound}. In addition, we know that given the current state of the network, $S_t$, the largest value that $A_{i}(t+1)$ can take is $A_{i}(t)+1$, i.e., there was not packet arrival from user $i$ to the receiver in the $t^\text{th}$ slot. Therefore, we can select $B=\sum\limits_{i=1}^{N} \frac{(A_{i}(t)+1)^2 + (A_{i}^{\text{max}})^2}{2} \geq \sum\limits_{i=1}^N \frac{\mathbb{E}\{A^2_{i}(t+1)|S_t\}+(A_{i}^\text{max})^2}{2}\text{.}$

\section{Proof of Theorem 2}
\begin{proof}
Suppose that a feasible policy $\omega$ exists, i.e., constraints \eqref{constropt: age} are satisfied. Suppose that, for the $\omega$ policy, the followings hold
	\begin{align}\nonumber
		&\mathbb{E}	\{(A_{i}^p(t)+1)(p_{i}s_{i}(t)+p_{i}\mu_{i}(t) - p_i s_i(t)\mu_{i}(t)) \\\nonumber
		&+(A_{i}(t)+1)(1-p_is_i(t)-\mu_{i}(t)p_i+p_i\mu_{i}(t)s_i(t))\} \\\label{bound}
		& \leq A_{i}^{\text{max}} - \epsilon_i\text{, } \forall i \in \mathcal{N}\text{,}\\\label{averagesub}
		&\mathbb{E} \{c^{*}(\epsilon)\} = c^{*} (\epsilon)\text{,}
	\end{align}
where $\epsilon>0$, $c^{*}$ is a suboptimal solution, and $\sum\limits_{i=1}^N\epsilon_{i} = \epsilon$. 

\noindent\eqref{ineq: dppbound} $\xRightarrow[]{\eqref{averagesub},\eqref{bound}}$ 
\begin{align}\nonumber
 \mathbb{E}\{L(\mathbf{X}(t+1))\} - \mathbb{E}\{L(\mathbf{X}(t))\} + V\mathbb{E}\{c(t)\}\leq\\ \label{ineq28}
 B - \epsilon \sum\limits_{i=1}^N\mathbb{E}\{X_{i}(t) \} + Vc^{*}(\epsilon)\text{,}
\end{align}
taking $\epsilon \rightarrow 0$, we obtain
\begin{align}\nonumber 
		\mathbb{E}\{c(t)\} \leq \frac{\mathbb{E}\{L(X_{i}(t)\}-\mathbb{E}\{L(X_i(t+1))\}}{V} + \frac{B}{V} + c^{*}\text{,}
\end{align}
taking the sum over $\tau= 0,\ldots,t-1$, we have
\begin{align}\nonumber
\frac{1}{t} \sum\limits_{\tau=0}^{t-1} \mathbb{E} \{c(\tau)\} & \leq \frac{-\mathbb{E}\{L(\mathbf{X}(t+1))\} + \mathbb{E}\{L(\mathbf{X}(0))\} Bt }{Vt} + c^{\text{opt}}\\
& \leq \frac{\mathbb{E}\{L(\mathbf{X}(0))\}}{Vt} + \frac{B}{V} + c^{\text{opt}}\text{,}
\end{align}
taking $t\rightarrow \infty$, we obtain
\begin{align}
		\lim_{t\rightarrow \infty} \sup\frac{1}{t} \sum\limits_{\tau=0}^{t-1}\mathbb{E}\{c(\tau)\} \leq c^{\text{opt}} + \frac{B}{V}\text{.}
\end{align}
That concludes the result of the second part of Theorem 1.

For proving stability of the virtual queues, we manipulate \eqref{ineq28} as
\begin{align}\nonumber
			\sum\limits_{i=1}^N\{X_{i}(t)\} & \leq \frac{B}{\epsilon} - \frac{\mathbb{E}\{L(\mathbf{X}(t+1))\}-\mathbb{E}\{L(\mathbf{X}(t))\}}{\epsilon}\\
			& - \frac{V\mathbb{E}\{c(t)\}}{\epsilon} + \frac{V}{\epsilon} c^{*}(\epsilon)\text{.}
\end{align}
By taking the sum over $\tau=0,\ldots,t-1$ and divide by $t$, we obtain
\begin{align}
\frac{1}{t}\sum\limits_{\tau=0}^{t-1}\sum\limits_{i=1}^{N}\mathbb{E} \{X_{i}(t)\} & \leq \frac{B}{\epsilon} - \frac{\mathbb{E}\{L(\mathbf{X}(t))\}-\mathbb{E}\{L(\mathbf{X}(0))\}}{t\epsilon}\\\nonumber
&-\frac{V\mathbb{E}\{c(t)\}}{\epsilon} + \frac{V}{\epsilon}c^{*}(\epsilon)\text{,}
\end{align}
neglecting the negative term and taking $t\rightarrow \infty$, we have 
\begin{align}\nonumber
		&\lim_{t\rightarrow \infty} \frac{1}{t} \sum_{\tau=0}^{t-1}\sum\limits_{i=1}^{N} \mathbb{E} \{X_{i}(\tau)\}  \leq \frac{B + V(-\mathbb{E}\{c(t\} + c^{*}(\epsilon)\})}{\epsilon}\text{,} \\\nonumber
\end{align}
considering that $\mathbb{E}\{c(t)\}\geq c^{\text{opt}}$, we get the final result as following 
\begin{align}
\lim_{t\rightarrow \infty} \sup\frac{1}{t} \sum_{\tau=0}^{t-1}\sum\limits_{i=1}^{N} \mathbb{E} \{X_{i}(\tau)\}  \leq \frac{B + V(  c^{*}(\epsilon)-c^{\text{opt}})}{\epsilon} \text{.}
\end{align}
This shows that the virtual queues $\{X_i\}_{i\in \mathcal{N}}$ are strongly stable.
\end{proof}
\end{appendices}
%\section{Acknowledgement}
%This work has been supported by the European
%Union's Horizon 2020 research and innovation programme under
%the Marie Sk\l{}odowska-Curie grant agreement No. $643002$.
\bibliography{MyBibAge}
\bibliographystyle{ieeetr}

\end{document}

%% file: acronyms.tex
\acrodef{AoI}{Age of Information}
\acrodef{IoT}{Internet of Things}
\acrodef{DPP}{Drift-Plus-Penalty}